\documentclass[12pt,twoside]{article}
\usepackage[utf8]{inputenc}

\usepackage{amsfonts}
\usepackage{amsmath}
\usepackage{amscd}
\usepackage{amsthm}
\usepackage{cite}
\usepackage{epsfig}
\usepackage{a4}
\usepackage[all,cmtip]{xy}

\newcommand{\I}{\mathrm{i}}        
\newcommand{\E}{\mathrm{e}}        

\newcommand{\Fun}{\operatorname{Fun}}
\newcommand{\SymFun}{\operatorname{SymFun}}

\newcommand{\Det}{\operatorname{det}}

\newcommand{\End}{\operatorname{End}}   

\newcommand{\tr}{\operatorname{tr}} 

\newcommand{\As}{\mathcal{A}}
\newcommand{\Bs}{\mathcal{B}}
\newcommand{\Cs}{\mathcal{C}}
\newcommand{\Ds}{\mathcal{D}}
\newcommand{\Ts}{\mathcal{T}}

\newcommand{\bvec}{\ensuremath{\mathbf b}}
\newcommand{\tvec}{\ensuremath{\mathbf t}}
\newcommand{\xvec}{\ensuremath{\mathbf x}}

\newcommand{\XXZ}{X\!X\!Z}
\newcommand{\XXX}{X\!X\!X}

\theoremstyle{plain}
\newtheorem*{example*}{Example}
\newtheorem*{conjecture*}{Conjecture}
\newtheorem{theorem}{Theorem}
\newtheorem*{lemma*}{Lemma}
\newtheorem*{remark}{Remark}

\theoremstyle{definition}

\newlength{\HFPP}       \HFPP5.4mm
\makeatletter
\@addtoreset{equation}{section}
\makeatother

\begin{document}

\thispagestyle{empty}

\begin{center}

{\Large {\bf Separation of Variables in the open XXX chain\\}}

\vspace{7mm}

{\large Holger Frahm\footnote[1]{e-mail: frahm@itp.uni-hannover.de}, 
Alexander Seel\footnote[2]{e-mail: alexander.seel@itp.uni-hannover.de}
and Tobias Wirth\footnote[3]{e-mail: tobias.wirth@itp.uni-hannover.de}}

\vspace{5mm}

Institut f\"ur Theoretische Physik, Leibniz Universit\"at Hannover,\\
Appelstr. 2, 30167 Hannover, Germany\\[2ex]

\vspace{20mm}

{\large {\bf Abstract}}

\end{center}

\begin{quote}
  We apply the Sklyanin method of separation of variables to the reflection
  algebra underlying the open spin-$\frac{1}{2}$ $\XXX$ chain with
  non-diagonal boundary fields.  The spectral problem can be formulated in
  terms of a $TQ$-equation which leads to the known Bethe equations for
  boundary parameters satisfying a constraint.  For generic boundary
  parameters we study the  asymptotic behaviour of the solutions of the
  $TQ$-equation.\\[2ex] 

\end{quote}

\clearpage

%
\section{Introduction}
%
Exact solutions of spin chain models have provided many insights into the
properties of interacting many-body systems subject to strong quantum
fluctuations.  Various methods have been established to study the spectrum and
nature of their low-lying excitations of such models as well as their
thermodynamical properties and even correlation functions without the need to
revert to perturbational approaches.
At the same time various problems concerning systems with open boundaries are
still not solved completely.  Even for the prototype spin-$\frac{1}{2}$ $\XXZ$
chain with general open boundary conditions techniques for the solution of the
spectral problem have been developed only recently
\cite{CaoX03,Nepo04,MeRM05,MuNS06,BaKo07,Galleas08}. 
This model, apart from being the simplest starting point for studies of
boundary effects in a correlated system, allows to investigate the approach to
a stationary state in one-dimensional diffusion problems for hard-core
particles \cite{GiEs05,GiEs06} and transport through one-dimensional quantum
systems \cite{CaWa08}.  Its Hamiltonian is given by
\begin{equation}\label{hamil}
\begin{aligned}
  \mathcal{H} = \sum_{j=1}^{L-1}&\Big[\sigma_j^x\sigma_{j+1}^x + \sigma_j^y\sigma_{j+1}^y +\big(\sigma_j^z\sigma_{j+1}^z - 1\big)\cosh\eta\Big] -\cosh\eta\\
+&\Big[\sigma_1^z \coth\xi^- + \frac{2\kappa^-}{\sinh\xi^-}\big(\sigma_1^x \cosh\theta^- + \I\sigma_1^y \sinh\theta^-\big)\Big]\sinh\eta\\
+&\Big[\sigma_L^z \coth\xi^+ + \frac{2\kappa^+}{\sinh\xi^+}\big(\sigma_L^x \cosh\theta^+ + \I\sigma_L^y \sinh\theta^+\big)\Big]\sinh\eta
\end{aligned}
\end{equation}
where $\sigma_j^\alpha$, $\alpha=x,y,z$ denote the usual Pauli matrices at site $j$ and the parameters $\xi^\pm$, $\kappa^\pm$, $\theta^\pm$ refer to the left and right boundary respectively.
The Hamiltonian (\ref{hamil}) is a member of a commuting family of operators
generated by a transfer matrix based on a representation of Sklyanin's
reflection algebra \cite{Skly88}.  While this establishes the integrability
the actual solution of the spectral problem through application of Bethe
ansatz methods has been impeded by the absence of a reference state, such as
the ferromagnetically polarized state with all spins up for the case of
diagonal boundary fields.
Interestingly, by imposing certain constraints obeyed by the left and right
boundary fields, the eigenvalues of the spin-$\frac{1}{2}$ $\XXZ$ chain and of
the isotropic spin-$S$ model can be obtained by means of the algebraic Bethe
ansatz \cite{CaoX03,MeRM05}.
In an alternative approach, Nepomechie \emph{et~al.} have been able to derive Bethe type equations whose roots parametrize
the eigenvalues of the Hamiltonian for special values of the anisotropy
\mbox{$\eta=\I\pi/(p+1)$} with $p$ a positive integer and where the transfer matrix obeys 
functional equations of finite
order \cite{MuNS06}. Their approach relies on the periodicity of the underlying
trigonometric $R$-matrix of the model which is missing in the rational limit
$\eta\to0$ of the isotropic chain. For generic values of the anisotropy the spectral problem has been formulated as a $TQ$-equation assuming that the large $j$-limit of the transfer matrices with spin $j$ in auxiliary space exists \cite{YaNeZh06}. No such constraints are needed in the derivation of a different set of recursion  relations for the diagonalization of (\ref{hamil}) based on the representation theory of the $q$-Onsager algebra\cite{BaKo07}. Very recently, Galleas has formulated another functional approach to determine the eigenvalues of (\ref{hamil}) in
the generic case without a reference state \cite{Galleas08}.  Taking certain
matrix elements of the transfer matrix involving the ferromagnetic pseudo
vacuum and the unknown eigenstates he derives the Bethe equations of
Nepomechie \emph{et al.} without the need to restrict the anisotropy to roots
of unity.  Unfortunately, this approach yields no information on the
eigenstates, therefore further studies of correlation functions are out of
reach for now.

In this paper we approach the problem by means of a different method which
circumvents the difficulties of the algebraic Bethe ansatz in the absence of a
reference state.  Within Sklyanin's functional Bethe ansatz (or separation of
variables method) \cite{Skly92} the eigenvalue problem is formulated using a
suitably chosen representation of the underlying Yang-Baxter algebra on a
space of certain functions. This approach was independently confirmed by use of Manin matrices \cite{ChFa07} and has proven its strength in
particular in models with non-compact target space lacking a reference state,
e.g.\ the Toda chain \cite{Skly85} or the sinh-Gordon model
\cite{ByTe06,Tesch07} where the spectral problem can be formulated in terms of
separable functional equations on this space which then have to be solved by
exploiting the analytical properties of the problem.

The article is organized as follows.
In sections \ref{IBC} and \ref{QD} we will review the basic properties of the
reflection algebra and its representation for the open $\XXZ$ spin chain. Then
in section \ref{FBA} we will implement the functional Bethe ansatz for the
restriction to the open $\XXX$ spin chain following Sklyanin's original
article \cite{Skly92}.  Finally in section \ref{TQ} we address the problem of
extracting the solution of the spectral problem from the resulting second
order difference equations and study the analytical properties of their
eigenfunctions. The last section is devoted to a concluding summary.


\section{Integrable Boundary Conditions} \label{IBC}
Sklyanin's construction \cite{Skly88} of integrable systems involving
boundaries is valid for a general class of integrable systems characterized by
an $R$-matrix of difference form $R(\lambda,\mu) =$ \mbox{$R(\lambda-\mu)$}$
\in \End(V\otimes V)$ ($V$ is a vector space with $\dim V  \in \mathbb{N}$ )
which not only satisfies the Yang-Baxter equation 
\begin{equation} \label{YBE}
R_{12}(\lambda-\mu)\,R_{13}(\lambda-\nu)\,R_{23}(\mu-\nu)
= R_{23}(\mu-\nu)\,R_{13}(\lambda-\nu)\,R_{12}(\lambda-\mu)
\end{equation}
but also several conditions such as symmetry with respect to the permutation
operator $P$ on $V \otimes V$ ($P\,x\otimes y = y \otimes x$), 
\begin{equation}
R(\lambda) = PR(\lambda)P\quad ,
\end{equation}
unitarity involving some complex function $\rho(\lambda)$,
\begin{equation}
R(\lambda)R(-\lambda) = \rho(\lambda)
\end{equation}
and crossing unitarity for another complex function $\widetilde\rho(\lambda)$,
\begin{equation}
R^{t_1}(\lambda)R^{t_1}(-\lambda-2\eta) = \widetilde\rho(\lambda) \quad.
\end{equation}
The parameter $\eta$ characterizes the $R$-matrix and the superscript $t_j$
denotes the transposition with respect to the $j$th space in the tensor
product $V\otimes V$. Here we will need the well-known $6$-vertex model
solution  
\begin{equation} \label{Rmatrix}
R(\lambda,\mu) =
  \begin{pmatrix}
     1 & 0 & 0 & 0 \\
     0 & b(\lambda,\mu) & c(\lambda,\mu) & 0 \\
     0 & c(\lambda,\mu) & b(\lambda,\mu) & 0 \\
     0 & 0 & 0 & 1
  \end{pmatrix} \qquad , \qquad
  \begin{gathered}
     b(\lambda,\mu) = \frac{\sinh(\lambda-\mu)}
                           {\sinh(\lambda-\mu+\eta)}\\
     c(\lambda,\mu) = \frac{\sinh \eta}{\sinh(\lambda-\mu+\eta)}
  \end{gathered}
\end{equation}
of the Yang-Baxter equation \eqref{YBE}. It generates the Hamiltonian of the
spin-$\frac12$ $\XXZ$ chain with 
\begin{equation}
\rho(\lambda) = 1 \quad ,\quad \widetilde\rho(\lambda) = \frac{\sinh\lambda\,\sinh(\lambda+2\eta)}{\sinh^2(\lambda+\eta)} \quad .
\end{equation}
Each solution $R(\lambda)$ of the Yang-Baxter equation fixes the structure
constants of a Yang-Baxter algebra  
\begin{equation}\label{YBA}
R_{12}(\lambda-\mu) T_1(\lambda) T_2(\mu) = T_2(\mu) T_1(\lambda)R_{12}(\lambda-\mu)
\end{equation}
with generators $T^\alpha_{\phantom{x}\beta}(\lambda)$, $\alpha,\beta=1,2$;
where $T_1(\lambda) = T(\lambda) \otimes I$, $T_2(\lambda) = I \otimes
T(\lambda)$ are the embeddings of the monodromy matrix $T(\lambda)$. 

Sklyanin's construction of open spin chains is based on the representations of
two algebras $\Ts^{(+)}$ and $\Ts^{(-)}$ defined by the relations
\begin{equation}\label{leftalg}
R_{12}(\lambda-\mu) \Ts_1^{(-)}(\lambda) R_{12}(\lambda+\mu)\Ts_2^{(-)}(\mu) 
=\Ts_2^{(-)}(\mu)R_{12}(\lambda+\mu)\Ts_1^{(-)}(\lambda)R_{12}(\lambda-\mu)
\end{equation}
\begin{multline}\label{rightalg}
R_{12}(-\lambda+\mu) \Ts_1^{(+)t_1}(\lambda) R_{12}(-\lambda-\mu-2\eta)\Ts_2^{(+)t_2}(\mu) =\\
=\Ts_2^{(+)t_2}(\mu)R_{12}(-\lambda-\mu-2\eta)\Ts_1^{(+)t_1}(\lambda)R_{12}(-\lambda+\mu) \quad  .
\end{multline}
We shall call $\Ts^{(+)}$ and $\Ts^{(-)}$ right and left reflection algebras
respectively. The transfer matrix  
\begin{equation}
t(\lambda) = \tr \Ts^{(+)}(\lambda)\Ts^{(-)}(\lambda)
\end{equation}
as a trace in auxiliary space is the central object under consideration
because it generates with $[t(\lambda),t(\mu)]=0$ a commuting family of
operators.

The explicit construction of integrable open boundary conditions for models
arising from the Yang-Baxter algebra starts with the $2\times2$ matrix  
\begin{equation}
\begin{aligned}
K(\lambda,\xi) &= \frac{1}{\sinh\xi \cosh\lambda}
\begin{pmatrix}
\sinh(\lambda+\xi) & \kappa \E^\theta \sinh(2\lambda) \\
\kappa \E^{-\theta} \sinh(2\lambda) & -\sinh(\lambda-\xi)
\end{pmatrix} 
 \\
  &=I_2 + \sigma^z \,\tanh\lambda \,\coth\xi 
+\frac{2\kappa\E^\theta\sinh\lambda}{\sinh\xi}\sigma^+
+\frac{2\kappa\E^{-\theta}\sinh\lambda}{\sinh\xi}\sigma^-
\end{aligned}
\end{equation}
originally found by de~Vega et al. \cite{VeGo93}.
It constitutes the known $c$-number representations $K^{(+)}(\lambda) =
\frac12 K(\lambda+\eta,\xi^+)$ and $K^{(-)}(\lambda) = K(\lambda,\xi^-)$ of the
reflection algebras with the obvious properties
\begin{equation}
\tr K(\lambda,\xi) = 2\,, \quad 
K^{(-)}(0) = I\,, \quad  \tr K^{(+)}(0) = 1 \,.
\end{equation}

The Hamiltonian \eqref{hamil} involving two neighbouring sites for the
interaction is connected, up to a factor, to the first derivative of
$t(\lambda)$ by looking at the expansion $t(\lambda) = 1 +
2\lambda\mathcal{H}+\ldots$ around the point $\lambda=0$.  Considering local
$L$-matrices building up the two representations
$T^{(+)}(\lambda)=L_L(\lambda)\cdots L_{M+1}(\lambda)$ and
$T^{(-)}(\lambda)=L_M(\lambda)\cdots L_1(\lambda)$ of \eqref{YBA} then by
construction
\begin{equation}
\begin{aligned}
\Ts^{(-)}(\lambda) &= T^{(-)}(\lambda) K^{(-)}(\lambda) T^{(-)-1}(-\lambda) \\
\Ts^{(+)t}(\lambda) &= T^{(+)t}(\lambda) K^{(+)t}(\lambda) \big(T^{(+)-1}\big)^t(-\lambda)
\end{aligned}
\end{equation}
are representations of the reflection algebras such that the normalized transfer matrix
\begin{equation}
t(\lambda) 
= K^{(+)}(\lambda) T(\lambda) K^{(-)}(\lambda)T^{-1}(-\lambda) \quad , \quad t(0)=1
\end{equation}
is independent of the factorization of $T(\lambda)=T^{(+)}(\lambda)T^{(-)}(\lambda)$. Thus we are free to choose
\begin{equation}
\Ts^{(+)}(\lambda) = K^{(+)}(\lambda) \quad , \quad 
\Ts^{(-)}(\lambda) = T(\lambda) K^{(-)}(\lambda) T^{-1}(-\lambda) \quad .
\end{equation}

In order to gain more symmetric arguments and to avoid inconvenient scalar
functions after applying the inversion formula 
\begin{equation}
T^{-1}(\lambda) = \frac{1}{(d_qT)(\lambda-\eta/2)} \sigma^y T^t(\lambda-\eta) \sigma^y
\end{equation} 
it is instructive to define the new object $U(\lambda+\eta/2) \equiv
\Ts^{(-)}(\lambda)\,(d_q T)(-\lambda-\eta/2)$ consisting of
\begin{equation}
\label{defU}
U(\lambda) = T(\lambda-\eta/2) K^{(-)}(\lambda-\eta/2) \sigma^y T^t(-\lambda-\eta/2) \sigma^y \quad .
\end{equation}
It is still a representation of the left reflection algebra with a $2\times2$
matrix in auxiliary space, 
\begin{equation}
U(\lambda) = 
\begin{pmatrix}
\As(\lambda) & \Bs(\lambda) \\
\Cs(\lambda) & \Ds(\lambda)
\end{pmatrix} \quad .
\end{equation}

The quantum determinant $(d_q T)(\lambda)$ is the central element (Casimir) of
the Yang-Baxter algebra (\ref{YBA}).  With the one-dimensional projector
$P_{12}^-$ onto the antisymmetric (singlet) state in the tensor product
$V\otimes V$ of auxiliary spaces the definition reads
\begin{equation}
\begin{aligned}
(d_q T)(\lambda) &= \tr_{12} P_{12}^- T_1(\lambda-\eta/2)T_2(\lambda+\eta/2) \\
 &= A(\lambda+\eta/2)\,D(\lambda-\eta/2) - B(\lambda+\eta/2)\,C(\lambda-\eta/2) \quad .
\end{aligned}
\end{equation}
Here, the trace $\tr_{12}$ is to be taken in both auxiliary spaces $1$ and $2$
of the tensor product $V\otimes V$ and the monodromy matrix $T$ enters with
the usual representation
\begin{equation}
T(\lambda) = 
\begin{pmatrix}
A(\lambda) & B(\lambda) \\
C(\lambda) & D(\lambda)
\end{pmatrix} \quad .
\end{equation}


\section{Quantum Determinants} \label{QD}
Like the Casimir for the Yang-Baxter algebra there exists a similar object for the left reflection algebra. It is defined as
\begin{equation}
(\Delta_q U)(\lambda) = \tr_{12} P_{12}^- U_1(\lambda-\eta/2) R_{12}(2\lambda-\eta)U_2(\lambda+\eta/2) \quad .
\end{equation}

To express $(\Delta_q U)(\lambda)$ in terms of the generators $\As(\lambda)$, $\Bs(\lambda)$, $\Cs(\lambda)$ and $\Ds(\lambda)$ it is instructive to use the combinations
\begin{equation}
\widetilde{\Ds}(\lambda) \equiv \sinh(2\lambda)\Ds(\lambda) -
\sinh\eta\,\As(\lambda) \,, \quad  
\widetilde{\Cs}(\lambda) \equiv \sinh(2\lambda+\eta) \Cs(\lambda)
\end{equation}
borrowed from the algebraic Bethe ansatz. Then the suggestive form of the quantum determinant reads 
\begin{equation}
(\Delta_q U)(\lambda) = \As(\lambda+\eta/2)\,\widetilde{\Ds}(\lambda-\eta/2) - \Bs(\lambda+\eta/2)\,\widetilde{\Cs}(\lambda-\eta/2) \quad .
\end{equation}

In case of the $c$-number representation $K(\lambda-\eta/2,\xi)$ for
$U(\lambda)$ connected to the left reflection algebra the relation 
\begin{equation}
  (\Delta_q K)(\lambda-\eta/2,\xi) =
  \frac{\sinh(2\lambda-2\eta)\cosh\lambda}{\cosh(\lambda-\eta)} \Det
  K(\lambda,\xi) 
\end{equation}
holds. Note that this relation is only valid for the shifted argument
$\lambda-\eta/2$ because the arising expressions are no longer of difference
form.

Appropriately transforming the boundary parameters $\xi$ and $\kappa$
according to the mapping \cite{Nepo04}
\begin{equation} \label{nepomap}
\sinh\alpha \cosh\beta = \frac{\sinh\xi}{2\kappa} \quad , \quad
\cosh\alpha\sinh\beta = \frac{\cosh\xi}{2\kappa} 
\end{equation}
the determinant $\Det K(\lambda,\xi)$ factorizes and its quantum version
decomposes to product form
\begin{equation}
  (\Delta_q K)(\lambda-\eta/2,\xi) = -\sinh(2\lambda-2\eta)
  \frac{\sinh(\lambda-\alpha)\cosh(\lambda-\beta)}{\sinh\alpha\cosh\beta\cosh(\lambda-\eta)}
  \frac{\sinh(\lambda+\alpha)\cosh(\lambda+\beta)}{\sinh\alpha\cosh\beta\cosh\lambda} \quad .
\end{equation}
It is obvious that in the parametrization (\ref{nepomap}) the model is
invariant under the simultaneous transformations $\alpha \to -\alpha$ and
$\beta\to\I\pi-\beta$.
%

As the quantum determinant respects co-multiplication, applying it to the full
representation (\ref{defU}) of the left reflection algebra with monodromy matrices
$T$ yields
\begin{equation} \label{qdetref}
(\Delta_q U)(\lambda) = (d_q T)(\lambda-\eta/2) \, (\Delta_q K)(\lambda-\eta/2,\xi) \,
(d_q T)(-\lambda-\eta/2) \quad .
\end{equation}

\begin{example*}
Consider the inhomogeneous periodic chain with inhomogeneities
$s_j\in\mathbb{C}$ at each lattice site $j=1 \ldots L$. Then the quantum
determinant of a fundamental $L$-matrix $L_j(\lambda)= R(\lambda-s_j)$ takes
the scalar value  
\begin{equation} \label{qdperiod}
(d_q L_j)(\lambda) = \sinh(\lambda-s_j+3\eta/2)\sinh(\lambda-s_j-\eta/2)\quad 
\end{equation}
yielding $(d_q T)(\lambda-\eta/2) = \Big[\prod_{j=1}^L
\sinh(\lambda-s_j+\eta)\sinh(\lambda-s_j-\eta)\Big]$ for a chain of $L$ local
spins $1/2$.
\end{example*}


\section{Functional Bethe Ansatz} 
\label{FBA} 
The entries of the monodromy matrix $U(\lambda)$ can be expressed in terms of
the operators $A(\lambda)$, $B(\lambda)$, $C(\lambda)$ and $D(\lambda)$ of the
periodic monodromy matrix $T(\lambda)$, i.e.
\begin{align} \notag
\Bs(\lambda) = &-\frac{\sinh(\lambda-\eta/2+\xi^-)}{\cosh(\lambda-\eta/2)\sinh\xi^-}
\frac{\sinh(2\lambda-\eta)}{\sinh(2\lambda)} B(-\lambda-\eta/2) \,A(\lambda-\eta/2) \\ \notag
&-\frac{\sinh(\lambda+\eta/2-\xi^-)}{\cosh(\lambda-\eta/2)\sinh\xi^-}
\frac{\sinh(2\lambda-\eta)}{\sinh(2\lambda)} B(\lambda-\eta/2) \,A(-\lambda-\eta/2)\\ \notag
&+ \frac{\kappa^- \E^{\theta^-}}{\sinh\xi^-}\frac{\sinh(2\lambda-\eta)}{\cosh(\lambda-\eta/2)}
A(\lambda-\eta/2)\,A(-\lambda-\eta/2) \\
&- \frac{\kappa^- \E^{\theta^-}}{\sinh\xi^-}\frac{\sinh(2\lambda-\eta)}{\cosh(\lambda-\eta/2)}
B(\lambda-\eta/2)\,B(-\lambda-\eta/2) \quad .
\end{align}
Obviously there is no easy pseudo vacuum $|0\rangle$ for an algebraic
Bethe ansatz to work ($\Bs(\lambda) |0\rangle = 0$) for non-diagonal boundaries
($\kappa^-\not=0$) in $K^{(-)}$.  Instead we can apply Sklyanin's functional
Bethe ansatz \cite{Skly92}.
 
Henceforth let us restrict to the rational case. Then we can choose the
boundary matrices diagonal in favour of a twisted monodromy matrix
$T(\lambda)$ as demonstrated in the following  
\begin{example*}
Applying the rational limit\footnote{\lq Rational limit\rq\ means to scale the
  arguments by a small parameter $\varepsilon \to 0$ and calling $\eta \to
  \varepsilon \I c$. Here only $\alpha$ is rescaled whereas $\beta$ remains
  unchanged. This immediately follows from \eqref{nepomap}.} to the
parametrization of the boundary matrix $K(\lambda,\xi)$ yields the similarity
transformation independent of the spectral parameter
\begin{equation}
K(\lambda,\xi) = M S
\left(\begin{matrix} 
\frac{\alpha+\lambda}{\alpha} & 0 \\ 0 & \frac{\alpha-\lambda}{\alpha}
\end{matrix}\right) \left(M S \right)^{-1}
\end{equation}
with the diagonal matrix to be a solution to the reflection algebra and the
$2\times2$ number matrices 
\begin{equation}
M =
\left(\begin{matrix}
\E^{+\theta/2} & 0 \\ 0 &\E^{-\theta/2}
\end{matrix}\right)  \quad , \quad 
S = \frac{1}{\sqrt{2\cosh\beta}}
\left(\begin{matrix}
\E^{\beta/2} & -\E^{-\beta/2} \\ \E^{-\beta/2} & \E^{\beta/2}
\end{matrix}\right) \quad .
\end{equation}
Thus in the transfer matrix $t(\lambda)=\tr K^{(+)}(\lambda)T(\lambda)
K^{(-)}(\lambda)T^{-1}(-\lambda)$ we are free to consider a diagonal outer
boundary matrix $K^{(+)}$ together with the $c$-number twist $(M^{(+)}
S^{(+)})^{-1}T(\lambda)$ of $T(\lambda)$.\footnote{The $gl(2)$ symmetry of the
  rational model allows to remove the twist of the monodromy matrix in favour
  of a twisted boundary matrix $\widetilde{K}^{(-)}= (M^{(+)}
  S^{(+)})^{-1}{K}^{(-)} M^{(+)} S^{(+)}$, see Ref.~\cite{MeRM05}.}
\end{example*}

\subsection*{Operator-Valued Zeroes}
The main goal of the functional Bethe ansatz is to treat the spectral problem
of the transfer matrix in a representation space of symmetric functions
manipulated by some shift operators, which descent from operator-valued zeroes
of the $\Bs$-operator.  
Starting from $\Bs(\lambda)$ in the rational limit reading
\begin{align} \notag
\Bs(\lambda) = -\frac{2\lambda-\I c}{\xi^-} \Bigg[ 
&\frac{\lambda+\xi^--\I c/2}{2\lambda} B(-\lambda-\I c/2) A(\lambda-\I c/2)\\ \notag
& + \frac{-\lambda+ \xi^- -\I c/2}{-2\lambda} B(\lambda-\I c/2) A(-\lambda-\I c/2) \\ \notag
& - \kappa^- \E^{\theta^-} A(\lambda-\I c/2)A(-\lambda -\I c/2)\\
& + \kappa^- \E^{-\theta^-} B(\lambda-\I c/2) B(-\lambda-\I c/2)\Bigg]
\end{align}
we observe the expression in the square brackets to be symmetric with respect
to $\lambda\to -\lambda$ and having no pole at $\lambda=0$. As the operators
$A$, $B$, $C$ and $D$ are polynomials\footnote{From here the
  $\varepsilon^L$-dependence of all periodic chain operators is suppressed.}
of degree $L$ with the known asymptotics 
\begin{equation}
\begin{aligned}
A &\sim \frac{ \exp ({\frac{\beta^+-\theta^+}{2}}) }{\sqrt{2\cosh\beta^+}}\lambda^L &&,& 
B &\sim \frac{ \exp ({\frac{\theta^+ - \beta^+}{2}}) }{\sqrt{2\cosh\beta^+}}\lambda^L\\
C &\sim -\frac{ \exp ({-\frac{\beta^+ + \theta^+}{2}}) }{\sqrt{2\cosh\beta^+}}\lambda^L &&,& 
D &\sim \frac{ \exp ({\frac{\beta^+ + \theta^+}{2}}) }{\sqrt{2\cosh\beta^+}}\lambda^L 
\end{aligned}
\end{equation}
we can factorize the square brackets in terms of $\lambda^2$. The asymptotic pre-factors arise from the twist $S^{-1}M^{-1}$ of the periodic monodromy
matrix $T(\lambda)$.

Thus $\Bs(\lambda)$ is polynomial with a simple zero at $\lambda=\I c /2$ and operator-valued coefficients assembling
\begin{equation} \label{Bopfac}
\Bs(\lambda) =
-\frac{2\lambda-\I c}{(-1)^L \xi^-} 
\frac{1-2\kappa^-\sinh(\theta^- - \theta^+ - \beta^+)}{2\cosh\beta^+}
\Big[\prod_{l=1}^L(\lambda^2-\widehat{x}_l^2)\Big] \quad .
\end{equation}
As $[\Bs(\lambda),\Bs(\mu)]$ = 0, according to the reflection algebra, we can
deduce $[\widehat{x}_j^2,\widehat{x}_k^2]=0$ for all $j,k=1\ldots L$ with the spectrum shown in the next
\begin{example*}
Setting $L=1$ in the explicit expression \eqref{Bopfac} for a spin $1/2$ representation with inhomogeneity $s_1$ yields for the argument $\lambda=0$ the form 
\begin{equation}
\widehat{x}_1^2 = 
\begin{pmatrix} (s_1+\I c/2)^2 & \\ & (s_1-\I c/2)^2
\end{pmatrix}
\end{equation}
of the operator-valued zero $\widehat{x}_1^2$ reading in a diagonalized
form. This fixes discrete sets $\Lambda_j \equiv \{s_j-\I c/2,s_j+ \I c/2\}$
representing the spectra of the coordinates $\widehat{x}_j$ except for a
global sign and resembling
$\mathbb{X}^L \equiv \Lambda_1\times\ldots\times\Lambda_L$. 
\end{example*}

For the forthcoming relations, to work with the simple zeroes $\widehat{x}_j$ instead of $\widehat{x}_j^2$, we refer to the supplemental 
\begin{remark}
Let $\widehat{x}_j$ be the operator-valued zeroes satisfying
$_{\lambda=\pm\widehat{x}_j}\left|\Bs(\lambda)\right.=0$. Then all
$\widehat{x}_j^2$ can be simultaneously diagonalized such that for all
$j,k=1\ldots L$ 
\begin{equation}
[\widehat{x}_j,\widehat{x}_k]=0 \quad ,\quad 
\widehat{x}_j =
\begin{pmatrix}
 s_j+\I c/2 & \\ & s_j-\I c/2
\end{pmatrix} \quad .
\end{equation}
\end{remark}



\subsection*{Conjugated Momenta}
With the operators $\widehat{x}_j$ the next problem is to calculate the
expression for the transfer matrix in the $\widehat{x}$-representation. For
this let us introduce first the \lq conjugated momenta\rq\ to the \lq coordinates\rq\
$\widehat{x}_j$. 

Considering $\As(\lambda)$ and $\widetilde{\Ds}(\lambda)$ as polynomials and
inserting the operator valued zeroes of $\Bs(\lambda)$ by \lq substitution from
the left\rq\ yields the new operators
\begin{equation}
\begin{aligned}
_{\lambda=\widehat x_j}\left|\As(\lambda)\right. = \sum_p \widehat{x}_j^p \As_p & \equiv X_j^-   \\
_{\lambda=\widehat x_j}\left|\widetilde{\Ds}(\lambda)\right. = \sum_p \widehat{x}_j^p \widetilde{\Ds}_p& \equiv X_j^+ \quad .
\end{aligned}
\end{equation}
Here $\As_p$ and $\widetilde{\Ds}_p$ denote operator-valued expansion
coefficients.  The commutation relations with the coordinates
$\widehat{x}_j$ are summarized in
\begin{theorem} \label{elementary}
Let $\widehat{x}_j$ be the operator-valued zeroes of $\Bs(\lambda)$ and $X_j^\pm$ their conjugated momenta related by the reflection algebra. Then
\begin{equation} \label{elementaryII}
X_j^\pm \widehat{x}_k = (\widehat{x}_k \pm \I c \delta_{jk}) X_j^\pm \quad .
\end{equation}
\end{theorem}
\begin{proof}
Consider the commutation relation 
\begin{align}
\nonumber \As(\lambda) \Bs(\mu) =& \frac{\sinh(\lambda+\mu-\eta)\sinh(\lambda-\mu-\eta)}{\sinh(\lambda+\mu)\sinh(\lambda-\mu)} \Bs(\mu) \As(\lambda) \\
                         &+\frac{\sinh\eta \sinh(2\mu-\eta)}{\sinh(2\mu)\sinh(\lambda+\mu)} \Bs(\lambda) \As(\mu) \\
\nonumber                &- \frac{\sinh \eta}{\sinh(2\mu) \sinh(\lambda+\mu)}\Bs(\lambda) \widetilde{\Ds}(\mu)
\end{align}
of $\As(\lambda)$ and $\Bs(\mu)$ and multiply it by $\sinh(\lambda+\mu)\sinh(\lambda-\mu)$. Then take its rational limit and insert the coordinates $\widehat{x}_j$ by \lq substitution from the left\rq. The expression reduces to
\begin{equation}
(\widehat{x}_j-\lambda)(\widehat{x}_j+\lambda ) X^-_j \Bs(\lambda ) = (\widehat{x} + \lambda - \I c)(\widehat{x}-\lambda-\I c) \Bs(\lambda) X^-_j \quad .
\end{equation}
Replacing $\Bs(\lambda)$ by its factorized form, cancelling the constant asymptotics
and multiplying by the inverse $(\widehat{x}_j^2-\lambda^2)^{-1}$ from the left the commutation relation
\begin{equation}
X_j^-\, \Big[\prod_{l=1}^L(\lambda^2-\widehat{x}_l^2)\Big] =
\Big[\lambda^2 - (\widehat{x}_j-\I c)^2\Big] \Big[\prod_{\overset{\scriptstyle{l=1}}{l\not=j}}^L(\lambda^2-\widehat{x}_l^2)\Big]\,X_j^-
\end{equation}
remains. Implying all expressions to be symmetric in $\widehat{x}_j$ to act on we arrive at the desired relation. 
Analogously the elementary commutation of $X_j^+$ with the coordinates arises from the commutation of $\Bs$ and $\widetilde{\Ds}$.
\end{proof}

The next natural step would be establishing the commutation relation between two $X^\pm$'s. However, it cannot be done directly because $X_j^\pm$ exceed per definition the representation space.


\subsection*{Representation space}
Following Sklyanin's approach the square brackets in operator \eqref{Bopfac} can be expanded into $\lambda^{2L} - \widehat{b}_1\lambda^{2(L-1)} \pm\ldots+\widehat{b}_L$ with commuting operators $\widehat{b}_j$ thus sharing a common system of eigenfunctions $f_\alpha$,
\begin{equation}
\widehat{b}_j f_\alpha = b_j^\alpha f_\alpha \quad , \quad \alpha=1 \ldots 2^L 
\end{equation}
where $2^L$ represents spin-$1/2$. To every point
$\bvec^\alpha=(b_1^\alpha,\ldots,b_L^\alpha) \in
\mathbb{B}^L\subset\mathbb{C}^L$ there corresponds only one eigenfunction
$f_\alpha$ and the representation space $W$ (e.g. for the $\XXX$ chain we have
$W=(\mathbb{C}^2)^{\otimes L}$) is isomorphic to the space $\Fun\mathbb{B}^L$.
\begin{example*}
  A possible realization of the eigenfunctions $f_\alpha$ is
\begin{equation}
(\widehat{b}_j f_\alpha)(\bvec^\beta) = b_j^\alpha f_\alpha(\bvec^\beta)
\end{equation}
where $\widehat{b}_j$ act as multiplication operators. Let
$\{\tvec_\alpha\in(\mathbb{C}^2)^{\otimes L}\big|(\tvec_\alpha)^\beta =
f_\alpha(\bvec^\beta)\}$ be a basis of $W$ then, with the constraint
$f_\alpha(\bvec^\beta) = \delta^\beta_\alpha$, it is indeed orthonormal and
complete.
\end{example*}

Since $\widehat{b}_n$ are the symmetric polynomials of the roots $\widehat{x}_j^2$ we are led to consider the mapping
\begin{equation}
\theta: \mathbb{C}^L \to \mathbb{C}^L\,  , \quad \xvec\mapsto\bvec
\end{equation}
given by the formula $b_n(\xvec)=s_n(\xvec)$. The $s_n(\xvec)$ are the
elementary symmetric polynomials of degree $n=1\ldots L$ of $c$-number variables
\begin{align}
s_1(\xvec) &= x_1^2 + x_2^2 + \ldots + x_L^2\notag\\ 
&\,\,\,\vdots\\
s_L(\xvec) &= x_1^2 x_2^2 \ldots x_L^2 \quad .\notag
\end{align}
The diagram
\begin{equation*}
{\xymatrix{
{\mathbb{X}^L} \ar[r]^\theta \ar@/^1.7pc/[rr]+/r1.4pc/+/u0.6pc/_g \ar@/_2pc/[rr]^{f \circ\, \theta} & \mathbb{B}^L \ar[r]^<<<<<f & \{0,1\} \subset \mathbb{C}
}}
\end{equation*}
of the combined mapping $f\circ\theta$ reveals the isomorphism between
$\Fun\mathbb{B}^L\cong W$ and the space of symmetric functions
$\SymFun\mathbb{X}^L$.  The set $\{0,1\}$ is the range of $f$ in the example
above. Thus the operator roots $\widehat{x}_j^2$ can be thought of
multiplication operators
\begin{equation} \label{multprop}
\widehat{x}_j^2 g(y_1,\ldots,y_L) = y_j^2 g(y_1,\ldots,y_L)
\end{equation}
in an extended representation space $\Fun\mathbb{X}^L\cong\widetilde{W}$ which is a non physical one. Recall all the results should only use the original space $\SymFun \mathbb{X}^L\cong W$, as the operators $\As$, $\Bs$, $\Cs$ and $\Ds$ map $\SymFun \mathbb{X}^L \to \SymFun \mathbb{X}^L$ or more sloppy $W \to W$. So for plainness we will use in the following the terms $W$ and $\widetilde{W}$ for the representation spaces instead.   

For the action of $X_j^\pm$ on a function $s\in \SymFun\mathbb{X}^L$
we need first to extend the operators from $W$ to $\widetilde{W}$ by the constant function
\begin{equation}
\omega(\xvec)=1 \text{ for all } \xvec\in\mathbb{X}^L \quad .
\end{equation}
Obviously $\omega$ is symmetric and thus belongs to the representation space $W$. Now define the action of $X_j^\pm$ on $\omega$ by
\begin{equation}
(X_j^\pm\omega)(\xvec) \equiv \Delta_j^\pm(\xvec) \quad .
\end{equation}
Then the functions $\Delta_j^\pm(\xvec)$ uniquely determine the action of $X_j^\pm$ on any vector $s$ which is identified due to the isomorphism with some symmetric function $s(x_1,\ldots,x_L)=(\widehat{s}\omega)(\xvec)$ created from the cyclic vector $\omega$ by the operator $\widehat{s}=s(\widehat{x}_1,\ldots,\widehat{x}_L)$. Thus
\begin{equation}
(X_j^\pm s)(\xvec) = (X_j^\pm \widehat{s} \omega)(\xvec) =s(E_j^\pm\xvec)(X_j^\pm\omega)(\xvec) = s(E_j^\pm\xvec)\Delta_j^\pm(\xvec) \quad .
\end{equation}
Here we introduced the shift operators
\begin{equation}
E_j^\pm: \mathbb{C}^L\to\mathbb{C}^L: (x_1,\ldots,x_j,\ldots,x_L) \mapsto (x_1,\ldots,x_j\pm\I c,\ldots,x_L)
\end{equation}
acting on some $L$-tuple of $c$-numbers. In the extended representation space $\widetilde{W}$ of not necessarily symmetric functions the action of $X_j^\pm$ then reads
\begin{equation}\label{keyrelation}
X_j^\pm = \Delta_j^\pm E_j^\pm
\end{equation}
with $\Delta_j^\pm=\Delta_j^\pm(\xvec)$. By the operator relation \eqref{keyrelation} we can now calculate the commutations of the momenta in
\begin{theorem} \label{theoremtwo}
Let $X_j^\pm$ be the conjugated momenta related to the coordinates $\widehat{x}_j$ by the reflection algebra. Then
\begin{equation}
\begin{aligned}
\left[X_j^\pm,X_k^\pm\right] &= 0\, \text{ for all } j,k=1\ldots L\\
\left[X_j^+,X_k^-\right] &= 0\, \text{ for all } j,k=1\ldots L \text{ but }
  j\not=k \quad .
\end{aligned}
\end{equation}
\end{theorem}
\begin{proof}
  Let us start with $X^-$ where the first assertion is obvious for $j=k$. Then
  it is enough to consider the cases $j=1$, $k=2$. Taking the rational limit
  of
\begin{equation}
\big[\As(\lambda),\As(\mu)\big] = \frac{\sinh\eta}{\sinh(\lambda+\mu)}\big[\Bs(\mu)\,\Cs(\lambda) - \Bs(\lambda)\,\Cs(\mu)\big] 
\end{equation}
and inserting $\lambda=\widehat{x}_1$ and $\mu=\widehat{x}_2$ by \lq substitution from the left\rq\ the RHS turns into zero and for the LHS we get
\begin{equation}
\begin{aligned}
_{\lambda=\widehat{x}_1,\mu=\widehat{x}_2}\left|\As(\lambda)\As(\mu)\right. 
&= \sum_{m,n}\widehat{x}_1^m\widehat{x}_2^n\As_m\As_n 
= \sum_{m,n}\widehat{x}_2^n\widehat{x}_1^m\As_m\As_n \\
&= \sum_{n}\widehat{x}_2^n X_1^- \As_n
= X_1^- \sum_n \widehat{x}_2^n\As_n
=X_1^- X_2^- \quad .
\end{aligned}
\end{equation}
In the same way starting from $\As(\mu)\As(\lambda)$ one obtains $X_2^- X_1^-$ and the
assertion is proven.  The commutation of $X^+$'s and mixed commutators
excluding the cases $j=k$ can be treated analogously by considering
\begin{equation}
\begin{aligned}
\left[\widetilde{\Ds}(\lambda),\widetilde{\Ds}(\mu)\right]
 &= -\sinh(2\lambda+\eta)\sinh(2\mu+\eta)\left[\As(\lambda),\As(\mu)\right]\\
\left[\widetilde{\Ds}(\lambda),\As(\mu)\right] 
 &= \frac{\sinh(\lambda+\mu)\sinh(2\lambda+\eta)}{\sinh(\lambda-\mu)}
 \left[\As(\lambda),\As(\mu)\right]
\end{aligned}
\end{equation}
in the rational limit.
\end{proof}

The remaining commutation involving the quantum determinant $\Delta_q$ is
summarized in 
\begin{theorem} \label{theoremthree}
Let $\widehat{x}_j$ and $X_j^\pm$  be the coordinates and conjugated
momenta related by the reflection algebra and $\Delta_q(\lambda)$ is the
quantum determinant. Then 
\begin{equation}
X_j^\pm X_j^\mp = \Delta_q(\widehat{x}_j\pm\I c/2) \text{ for all }
j,k=1\ldots L \quad .
\end{equation}
\end{theorem}
\begin{proof}
Substituting the operator-valued zeroes $\widehat{x}_j$ into the quantum
determinant \eqref{qdetref} one obtains 
\begin{equation}
\begin{aligned}
\Delta_q(\widehat{x}_j-\I c/2) &= \sum_{m,n} \widehat{x}_j^m (\widehat{x}_j-\I c)^n \As_m\widetilde{\Ds}_n = \sum_{m,n} (\widehat{x}_j-\I c)^n \Big(\widehat{x}_j^m \As_m\Big)\widetilde{\Ds}_n \\
&= \sum_n (\widehat{x}_j-\I c)^n X_j^- \widetilde{\Ds}_n = X_j^- \sum_n \widehat{x}_j^n\widetilde{\Ds}_n\\
&= X_j^-X_j^+
\end{aligned}
\end{equation}
and analogously  $\Delta_q(\widehat{x}_j+\I c/2)=X_j^+ X_j^-$ exerting the
reflection algebra. 
\end{proof}

\begin{remark}
  The remaining zero $\I c/2$ of $\Bs(\lambda)$ is an exception and renders
  the operators $\As(\I c/2) = d_q(-\I c/2)$ and $\widetilde{\Ds}(\I c/2) = 0$
  to be constant yielding $\Delta_q(\I c) = 0$.
\end{remark}
%


\subsection*{Representation of $\Delta^\pm$}
Applying $X_j^\pm X_j^\mp = \Delta^\pm_j E_j^\pm \Delta_j^\mp E_j^\mp$ to an arbitrary function $g\in\widetilde{W}$ induces the sequence
\begin{equation}
\begin{aligned}
(X_j^\pm X_j^\mp g)(\xvec) &= (\Delta^\pm_j E_j^\pm \Delta_j^\mp E_j^\mp g)(\xvec)
        = \Delta_j^\pm(\xvec)(E_j^\pm \Delta_j^\mp E_j^\mp g)(\xvec) \\
&= \Delta_j^\pm(\xvec)(\Delta_j^\mp E_j^\mp g)(E_j^\pm\xvec)
= \Delta_j^\pm(\xvec)\Delta_j^\mp(E_j^\pm\xvec)( E_j^\mp g)(E_j^\pm\xvec)\\
&= \Delta_j^\pm(\xvec)\Delta_j^\mp(E_j^\pm\xvec) g(\xvec) \\
&\overset{!}{=} \Delta_q(x_j\pm\I c/2) g(\xvec)
\end{aligned}
\end{equation}
relating the representations $\Delta_j^\pm$ to the quantum determinant $\Delta_q$.
In the case of a finite dimensional representation of the generators
$\{\widehat{x}_j,X_j^\pm\}_{j=1}^L$ such that the spectrum $\mathbb{X}^L$
shows no multiple points the problem of constructing such a representation is
equivalent to that of determining the functions $\{\Delta_j^\pm\}_{j=1}^L$ on
$\mathbb{X}^L$ satisfying 
\begin{equation}
\begin{aligned}
\Delta_m^\pm(\xvec)\Delta_n^\pm(E_m^\pm\xvec) &= \Delta_n^\pm(\xvec)\Delta_m^\pm(E_n^\pm\xvec)  &&\text{for all } n,m\\
\Delta_m^+ (\xvec)\Delta_n^- (E_m^+ \xvec) &= \Delta_n^- (\xvec)\Delta_m^+ (E_n^-\xvec)   &&\text{for all } n,m \quad \text{but } n\not=m\\
\Delta_q(\widehat{x}_j\pm\I c/2) &= \Delta_j^\pm(\xvec)\Delta_j^\mp(E_j^\pm\xvec) &&\text{for all } j
\end{aligned}
\end{equation}
arising from theorems \ref{theoremtwo} and \ref{theoremthree}. The above
relations are not defined when the shifts $E_j^\pm$ move the point $\xvec$ out
of $\mathbb{X}^L=\Lambda_1\times\ldots\times\Lambda_L$. 
This means $\{\Delta_j^\pm\}_{j=1}^L$ have to vanish on the boundary 
\begin{equation}
\partial\mathbb{X}_j^\pm \equiv  \{\xvec\in\mathbb{X}^L | E_j^\pm\xvec \in \mathbb{C}^L\backslash\mathbb{X}^L\}
\end{equation}
 of the set $\mathbb{X}^L$. For the open $\XXX$ chain with $\Lambda_j=\{s_j-\I c/2,s_j+\I c /2\}$ this is clear from the following
\begin{example*}
  The vanishing of $\Delta_j^\pm(\xvec)$ on the boundary
  $\partial\mathbb{X}_j^\pm$ can be directly seen from the explicit
  factorization of $\Delta_q(\lambda) =
  \Delta^-(\lambda+\eta/2)\Delta^+(\lambda-\eta/2)$ into

\begin{equation}
\label{deltapm}
\begin{aligned}
\Delta^-(\lambda) &= \frac{\lambda-\I c/2+\alpha^-}{(-1)^L\alpha^-} 
\Big[\prod_{l=1}^L (\lambda-s_l+\I c/2)(\lambda+s_l+\I c/2\Big] \\
\Delta^+(\lambda) &= -(2\lambda-\I c)\varepsilon\frac{\lambda+\I c/2-\alpha^-}{(-1)^L\alpha^-} 
\Big[\prod_{l=1}^L (\lambda-s_l-\I c/2)(\lambda+s_l-\I c/2\Big] 
\end{aligned}
\end{equation}

considered in the rational limit indicated by $\varepsilon\to0$.
\end{example*}

\subsection*{Spectral Analysis}
Now let us return to the original problem, the spectral analysis of the new
transfer matrix 
\begin{math}
\tau(\lambda) \equiv \tr K(\lambda+\eta/2,\xi^+)U(\lambda)/{2}
\end{math}
in the rational limit
\begin{align} \notag
\tau(\lambda) &= \frac{(\lambda+\I c/2)(\lambda+\xi^+-\I c/2)}{2\lambda\xi^+} \As(\lambda)
-\frac{1}{\varepsilon}\frac{\lambda-\xi^++\I c/2}{4\lambda\xi^+} \widetilde{\Ds}(\lambda)\\
&\phantom{=}+ \frac{(\lambda+\I c/2)\kappa^+}{\xi^+} \Big[\E^{\theta^+} \Cs(\lambda) + \E^{-\theta^+}    \Bs(\lambda)\Big] 
\end{align}
and mind the scaling factor $\varepsilon\to0$. To plug in the zeroes
$\widehat{x}_j$ by \lq substitution from the left\rq\ we have to get rid of
$\Cs(\lambda)$ by diagonalizing $K^{(+)}$. Thus only the first line  
remains. The diagonalization does not change the quantum determinant
$\Delta_q(\lambda)$ and the eigenvalue problem
$\tau(\lambda) \varphi = \Lambda(\lambda)\varphi$ can be solved by
\lq substitution from the left\rq\ reading 
\begin{equation} \label{halfTQ}
_{\lambda=\widehat{x}_j}\left|\tau(\lambda)\right. = \frac{(\widehat{x}_j+\I c/2)(\widehat{x}_j+\alpha^+-\I c/2)}{2\widehat{x}_j\alpha^+} X_j^-
-\frac{1}{\varepsilon}\frac{\widehat{x}_j-\alpha^+ + \I c/2}{4\widehat{x}_j\alpha^+} X_j^+ \quad .
\end{equation} 
With this representation at hand one observes \lq separation of variables\rq\
suggesting the product ansatz 
\begin{equation}
\varphi = \Big[\prod_{l=1}^L Q(x_l)\Big]
\end{equation}
for the eigenfunction $\varphi\in\SymFun \mathbb{X}^L \cong W$ symmetric in
its arguments $x_l$. To explicitly apply the operator-valued expression \eqref{halfTQ} one should clarify its behaviour by the following

\begin{lemma*}
The action of the combined expression $\widehat{x}_j X_j^\pm$ by \lq substitution from the left\rq\ onto a symmetric function $s=s(x_1,\ldots,x_L)$ is given by
\begin{equation} \label{conjecture2}
\widehat{x}_j X_j^\pm s(\xvec) = (\widehat{x}_j X_j^\pm s)(\xvec) = x_j (X_j^\pm s)(\xvec) = x_j \,\Delta^\pm_j(\xvec) s(E_j^\pm\xvec) \quad .
\end{equation}
\end{lemma*}

Then applying \eqref{halfTQ} to $\varphi$ only the $j$th
argument is affected such that the problem separates and
\begin{equation}
\begin{aligned}
\label{fullTQ}
\Lambda({x}_j) Q({x_j}) =& \frac{({x}_j+\I c/2)({x}_j+\alpha^+-\I
  c/2)}{2{x}_j\alpha^+}\Delta^-(x_j)Q(x_j - \I c) \\ 
 &\qquad -\frac{1}{\varepsilon}\frac{{x}_j-\alpha^+ + \I c/2}{4{x}_j\alpha^+}
 \Delta^+(x_j) Q(x_j + \I c)  
\end{aligned}
\end{equation}
holds. Here we used \eqref{conjecture2} with the allowed arguments $x_j \in \Lambda_j = \{s_j-\I c/2,s_j+\I c/2\}$ on the grid entering $\Delta_j^\pm(\xvec) = \Delta^\pm(x_j)$.  \\

\begin{remark}
Along with the normalization $(d_qT)(-\lambda)$, c.f. \eqref{qdperiod}, and a shift in the arguments
the original transfer matrix $t(\lambda)$ is related to $\tau(\lambda)$ via 
\begin{math}
\tau(\lambda) = (d_qT)(-\lambda)\, t(\lambda-\eta/2)
\end{math}.

\end{remark}


\section{TQ-Equations} \label{TQ}
The eigenvalue problem as formulated in Eq.~(\ref{fullTQ}) reduces to a system
of homogeneous linear equations due to the fact that $\Delta^\pm(x_j^\pm)=0$
at the points $x_j^\pm = s_j\pm\I c/2$:
\begin{equation}
\begin{aligned}
\Lambda(x_j^+) Q(x_j^+) &= \frac{(x_j^++\I c/2)(x_j^++\alpha^+-\I c/2)}{2 x_j^+ \alpha^+} \Delta^-(x_j^+) Q(x_j^-) \\
\Lambda(x_j^-) Q(x_j^-) &= -\frac{1}{\varepsilon} \frac{x_j^- - \alpha^+ + \I c/2}{4 x_j^- \alpha^+} \Delta^+(x_j^-) Q(x_j^+) \quad .
\end{aligned}
\end{equation}
For pairwise different inhomogeneities, $s_j\ne s_k$ for $j\ne k$, these
linear equations allow for a non-trivial solution provided that the following
functional equation for the eigenvalues $\Lambda$ are satisfied\footnote{If
  $n$ of the inhomogeneities coincide the $(n-1)$ derivatives of this 
equation at this value of $s_j$ have to be taken into account in addition.}
\begin{equation} \label{funkequ}
\Lambda(s_j+\I c/2)\Lambda(s_j - \I c/2) = -\frac{s_j+\I c}{2\varepsilon}\frac{s_j-\alpha^+}{(2s_j-\I c)\alpha^+}
\frac{s_j+\alpha^+}{(2s_j+\I c)\alpha^+} \Delta_q(s_j)\,,\quad j=1\ldots L.
\end{equation} 
Using the known asymptotic form of the even polynomial $\Lambda(\lambda) =
\Lambda(-\lambda)$ we are led to  the ansatz
\begin{equation}
\Lambda(\lambda) = \frac{(-1)^L}{\alpha^+\alpha^-}
\frac{\sinh\beta^+\sinh\beta^- + \cosh(\theta^+ - \theta^-)}{\cosh\beta^+\cosh\beta^-}\lambda^{2L+2} +
a_{2L}\lambda^{2L} + a_{2L-2}\lambda^{2L-2} + \ldots + a_0\,.
\end{equation}
The $(L+1)$ unknown coefficients $a_j$ are determined by Eqs.~(\ref{funkequ})
and the constraint $\Lambda(\I c/2) = d_q(-\I c/2)$ with the quantum
determinant $d_q(\lambda)$ of the periodic chain.
Thus the solution of the spectral problem amounts to finding the common roots
$\{a_{2j}^{(\nu)}\}_{j=0}^L$, $\nu=1\ldots2^L$, of these polynomial equations.
This task is of the same complexity as finding the eigenvalues of the spin
chain Hamiltonian directly and therefore this approach is limited to small
system sizes where we have checked numerically that it does indeed
yield the complete spectrum.


To compute the eigenvalue of the transfer matrix or the spin chain Hamiltonian
in the thermodynamic limit $L\to\infty$ the functional equations introduced
above need to be analyzed beyond the set $\mathbb{X}^L$ using explicitly the
analytic properties of the functions therein.

\begin{remark}
  Note that the functional equations (\ref{funkequ}) hold at the discrete
  points $s_j$ only: treating the $s_j$ as a continuous variable and applying
  standard Fourier techniques one can compute $\ln\Lambda(\lambda)$ and
  thereby the corresponding eigenvalue of the spin chain Hamiltonian
  (\ref{hamil}), i.e.\ $\mathcal{H} = \I c \,(\partial_\lambda\ln
  \tau)(\I c/2) = \I c \,t^\prime(0)$ in the homogeneous limit.  For 
  $|\alpha^\pm| > c/2$ one obtains
  \begin{equation}
  \begin{aligned}
{\I c}\frac{\partial\ln\Lambda}{\partial\lambda}(\I c/2) =&
\psi({|\alpha^+|}/{2c}) - \psi({|\alpha^+|}/{2c}+1/2) + {c}/{|\alpha^+|} \\ 
&+\psi({|\alpha^-|}/{2c}) - \psi({|\alpha^-|}/{2c}+1/2) +
{c}/{|\alpha^-|} \\[.5ex]
&+ \pi - 2\ln2 -1 + (2-4\ln2)L
  \end{aligned}
  \end{equation}
  which is for imaginary $\alpha^\pm$ the known energy eigenvalue of the
  $\XXX$ spin chain with diagonal boundary fields \cite{FrZv97b} ($\psi$ is
  the digamma function). 
  However, the non-diagonal contributions and corrections of the order $1/L$
  are not included.  This is a consequence of neglecting the corrections to
  Eqs.~(\ref{funkequ}) away from the points $s_j$.
\end{remark}

Instead we go back one step and consider the Eqs.~\eqref{fullTQ} for general
arguments $x_j\to\lambda$.  Formally, this is a second order difference
equation reminiscent of Baxter's $TQ$-equation \cite{baxter:book}.  The
analysis above leading to Eqs.~(\ref{funkequ}) has been based on the singular
points of the $TQ$-equation at the boundaries $\partial\mathbb{X}^L$, i.e.\
points where one of the coefficients $\Delta^\pm$ vanishes.  Away from these
points there exist two independent solutions to (\ref{fullTQ}) and one needs
some information on the properties of the unknown functions $Q(\lambda)$ in
this formulation of the spectral problem which has to be solved for polynomial
eigenvalues $\Lambda(\lambda)$ of the transfer matrix.

In cases where a pseudo vacuum exists and the algebraic Bethe ansatz is
applicable to the solution of the problem the $Q$-functions are known to be
symmetric polynomials $Q(\lambda) = \prod_{\ell=1}^M(\lambda -
v_\ell)(\lambda + v_\ell)$ with roots $v_\ell$ satisfying Bethe ansatz
equations.  Note that only in these cases the constant function $\omega=1$
introduced in the construction of the representation of the Yang-Baxter
algebra on the space $\widetilde W$ can be identified with the pseudo vacuum
$|0\rangle$.

In general, the $TQ$ equation can be rewritten as  a recursion relation
\begin{equation}
Q(\lambda+\I c) = a(\lambda) Q(\lambda) + b(\lambda) Q(\lambda-\I c)
\end{equation}
for the function $Q(\lambda)$ or equivalently, with the auxiliary function
$P(\lambda + \I c) \equiv Q(\lambda)$,
\begin{equation} \label{recursion}
\begin{pmatrix} Q(\lambda+\I c) \\ P(\lambda + \I c)
\end{pmatrix} =
\begin{pmatrix} a(\lambda) & b(\lambda) \\ 1 & 0
\end{pmatrix}
\begin{pmatrix} Q(\lambda) \\ P(\lambda)
\end{pmatrix} \quad .
\end{equation}
The coefficients $a(\lambda)$ and $b(\lambda)$ are obtained from the $TQ$-equation (\ref{fullTQ}) and show
constant asymptotics for large values of their arguments
\begin{equation}
\begin{aligned}
a(\lambda) &= -\frac{\Lambda(\lambda)}{\Delta^+(\lambda)}\frac{4\varepsilon\lambda\alpha^+}{\lambda-\alpha^++\I c/2} \sim 2\frac{\sinh\beta^+\sinh\beta^-+\cosh(\theta^+-\theta^-)}{\cosh\beta^+\cosh\beta^-}\,,\\
b(\lambda) &= \frac{\Delta^-(\lambda)}{\Delta^+(\lambda)}\frac{2(\lambda+\I c/2)(\lambda+\alpha^+-\I c/2)\varepsilon}{\lambda-\alpha^++\I c/2} \sim -1\,.
\end{aligned}
\end{equation}
This allows to solve the recursion relations in the asymptotic regime
$|\lambda | \gg 1$ yielding
\begin{equation}
Q(\lambda + \I nc) = \frac{\lambda_1^n-\lambda_2^n}{\lambda_1-\lambda_2} Q(\lambda+\I c) -
\lambda_1\lambda_2\frac{\lambda_1^{n-1}-\lambda_2^{n-1}}{\lambda_1-\lambda_2} Q(\lambda) \quad .
\end{equation}
Here $n$ is an integer and $\lambda_{1,2} = \E^{\pm\phi}$ are the eigenvalues
of the asymptotical matrix of coefficients in (\ref{recursion}), $\cosh\phi =
({\sinh\beta^+\sinh\beta^-+\cosh(\theta^+-\theta^-)})/({\cosh\beta^+\cosh\beta^-})$. 
Ordering the eigenvalues as $|\lambda_1|>|\lambda_2|$ we obtain for fixed
$\lambda$ and large $n$ the leading term $Q(\lambda+\I nc) \sim\lambda_1^n$
suggesting the following ansatz for the asymptotic form 
\begin{equation} \label{asymp}
Q(\lambda) \sim \exp\big(\frac{\lambda \phi}{\I c}\big) \times \text{polynomial
  in } \lambda \quad . 
\end{equation}
Here the polynomial form of the subleading part assures that the eigenvalue
$\Lambda(\lambda)$ of the transfer matrix remains polynomial.  Note, that
since $\Lambda(\lambda)$ is an even function of its argument there exists
always a second solution $Q(-\lambda)$ to the $TQ$ equation which decays
exponentially for $\lambda\to\infty$.

Only in two cases, namely $\phi=0$ and $\I\pi$ or, equivalently,
\begin{equation}
\label{constraintRL}
\cosh(\theta^+-\theta^-) = \pm\cosh(\beta^+\mp\beta^-)\,,
\end{equation}
the exponential factor disappears and the $TQ$ equation can be solved by an
even polynomial: in the first case Eq.~(\ref{asymp}) implies that
$Q(\lambda)=\prod_{\ell=1}^{M^{(+)}} (\lambda-v_\ell)(\lambda+v_\ell)$.  For
$\phi=\I\pi$, the exponential factors can be removed by the transformation
$Q(\lambda)=\exp(\I\lambda\pi/\I c)Q^\prime(\lambda)$ resulting in a $TQ$
equation for $Q^\prime$:
\begin{equation}
\begin{aligned}
\Lambda({x}_j) Q^\prime({x_j}) =& -\frac{({x}_j+\I c/2)({x}_j+\alpha^+-\I
  c/2)}{2{x}_j\alpha^+}\Delta^-(x_j)Q^\prime(x_j - \I c) \\ 
 &\qquad +\frac{1}{\varepsilon}\frac{{x}_j-\alpha^+ + \I c/2}{4{x}_j\alpha^+}
 \Delta^+(x_j) Q^\prime(x_j + \I c) \,.
\end{aligned}
\end{equation}
Again, it follows from the asymptotic analysis that this equation allows for a
polynomial solution $Q^\prime(\lambda) = \prod_{\ell=1}^{M^{(-)}}
(\lambda-v_\ell) (\lambda+v_\ell)$ whose existence has been verified by
numerical analysis for small systems.

In both cases the spectrum is determined by the roots of these polynomials.
To guarantee analyticity of the transfer matrix eigenvalues $\Lambda(\lambda)$
the $v_j$, $j=1\ldots M^{(\pm)}$, have to satisfy the Bethe ansatz equations
\begin{equation}
\label{bae_poly}
\begin{aligned}
  \frac{v_j+\alpha^--{\I c}/{2}}{v_j-\alpha^-+{\I c}/{2}} \,
  \frac{v_j+\alpha^+-{\I c}/{2}}{v_j-\alpha^++{\I c}/{2}} 
 \bigg[ \prod_{l=1}^L &
    \frac{v_j-s_l+{\I c}/{2}}{v_j-s_l-{\I c}/{2}} \,
    \frac{v_j+s_l+{\I c}/{2}}{v_j+s_l-{\I c}/{2}}\bigg] =\\
 & \phantom{xxx} = \bigg[\prod_{\overset{\scriptstyle{k=1}}{k\not=j}}^{M^{(\pm)}}
    \frac{v_j-v_k+\I c}{v_j-v_k-\I c}\,
    \frac{v_j+v_k+\I c}{v_j+v_k-\I c}\,\bigg] \quad .
\end{aligned}
\end{equation}

Note that Eq.~(\ref{constraintRL}) is equivalent to the constraint that the
boundary matrices $K^{(\pm)}$ can be simultaneously diagonalized or brought to
triangular form.  In this case Eqs.~(\ref{bae_poly}) can be obtained by means
of the algebraic Bethe ansatz \cite{MeRM05} or in the rational limit from the
$TQ$-equation approach for the open $\XXZ$ chain \cite{YaNeZh06}.  In this
trigonometric case the complete set of eigenvalues is obtained from two sets
of Bethe equations which both reduce to (\ref{bae_poly}).  This is due to the
invariance of the model under the change of parameters $\alpha \to -\alpha$
and $\beta\to\I\pi-\beta$ which maps $\phi=0 \leftrightarrow \I\pi$, see
Eq.~(\ref{nepomap}).
As another difference to the situation in the $\XXZ$ model the number of Bethe
roots is not restricted by the constraint on the boundary fields
and we have to consider solutions of the $TQ$-equations \rq beyond the
equator\rq, $M^{(\pm)}>L/2$.

\section{Summary}
In this paper we have extended Sklyanin's functional Bethe ansatz method to
the open $\XXX$ chain with non-diagonal boundary fields.  Within this
framework we have derived a single $TQ$-equation (\ref{fullTQ}) which
determines the spectrum of this model for any values of the boundary
parameters.  This supports the approach of Yang \textit{et al.}
\cite{YaNeZh06} who obtain a $TQ$-equation for the $\XXZ$ model assuming the
existence of a certain limit in the auxiliary space.

The $TQ$-equation allows for a solution in terms of polynomials for the
function $Q$ provided that a constraint between the left and right boundary
field is satisfied.  In this case the solution is parametrized by the roots of
one set of Bethe ansatz equations.  In our numerical study of small systems we
obtain the complete spectrum from these equations, when solutions \lq beyond
the equator\rq, i.e.\ polynomial $Q$ with degree $>L/2$, are taken into
account.  This is in contrast to the trigonometric case where two types of
Bethe equations are required and the degrees of the corresponding
$Q$-functions are fixed by the constraint.  We suppose that this feature of
the $\XXX$ case is a consequence of the rational limit.

If the constraint between the boundary fields is missing only the asymptotic
(exponential) behaviour of the $Q$-functions is obtained from the $TQ$
equation, the subleading terms have to be chosen such that the eigenvalues of
the transfer matrix remain polynomial.  In Sklyanin's approach the
$Q$-functions contain all the information on the eigenstates of the model.
Therefore, their determination for generic boundary parameters is necessary to
tackle the problem of computing norms and scalar products within this
approach.  To make progress in this direction it should be useful to
investigate how the recent construction of Galleas \cite{Galleas08} connects
to the $TQ$-equation (\ref{fullTQ}).  His solution of the spectral problem for
$\XXZ$ chains with non-diagonal boundaries is given in terms of the zeroes of
the transfer matrix eigenvalues and two complementary sets of numbers which
parametrize matrix elements of certain elements of the Yang-Baxter algebra.
They satisfy equations reminiscent of the nested Bethe ansatz used to solve
models of higher-rank symmetries.  Further studies are necessary to see
whether this parametrization of the spectrum in terms of $\mathcal{O}(L)$
complex numbers can be used to obtain a closed expression for generic
(non-polynomial) $Q$-functions.  This would be of great importance for the
applicability of the $TQ$-equation to solve the spectral problem of integrable
quantum chains.\\

{\bf Acknowledgements.}  The authors would like to thank A.~Kl\"umper,
J.~Teschner and S.~Niekamp for helpful discussions. 
This work has been supported by the Deutsche Forschungsgemeinschaft under
grant numbers Se~1742/1-1 and FR~737/6.

\providecommand{\bysame}{\leavevmode\hbox to3em{\hrulefill}\thinspace}
\providecommand{\MR}{\relax\ifhmode\unskip\space\fi MR }
\providecommand{\MRhref}[2]{%
  \href{http://www.ams.org/mathscinet-getitem?mr=#1}{#2}
}
\providecommand{\href}[2]{#2}


\end{document}